\documentclass[letter,twocolumn]{jpsj3}
\usepackage{txfonts}
\usepackage{here}
\usepackage{color}

\title{Evidence for a New Magnetoelectric Effect of Current-Induced Magnetization in a Toroidal Magnetic Ordered State of UNi$_{4}$B}

\author{Hiraku \surname{Saito}$^1$\thanks{E-mail:h3110@phys.sci.hokudai.ac.jp}, Kenta \surname{Uenishi}$^1$, Naoyuki \surname{Miura}$^1$, Chihiro \surname{Tabata}$^1$\thanks{Present address: Condensed Matter Research Center and Photon Factory, Institute of Materials Structure Science, High Energy Accelerator Research Organization, Tsukuba, Ibaraki 305-0801, Japan}, \\
Hiroyuki \surname{Hidaka}$^1$, Tatsuya \surname{Yanagisawa}$^1$, and Hiroshi \surname{Amitsuka}$^1$}
\inst{$^1$Graduate School of Science, Hokkaido University, Sapporo 060-0810, Japan}

\abst{

Magnetization measurements under direct electric currents were performed for toroidal magnetic ordered state of UNi$_4$B to test a recent theoretical prediction of current-induced magnetization in a metallic system lacking local inversion symmetry.
We found that each of the electric currents parallel to [2$\bar{1}$$\bar{1}$0] and [0001] in the hexagonal 4-index notation induces uniform magnetization in the direction of [01$\bar{1}$0].
The observed behavior of the induced magnetization is essentially consistent with the theoretical prediction; however it also shows an inconsistency suggesting that  the antiferromagnetic state of UNi$_4$B could not simply be regarded as a uniform toroidal order in the ideal honeycomb  layered  structure.

}
\begin{document}
\maketitle


The behavior of solids without space-inversion symmetry is one of the most attractive topics in the modern condensed matter physics in the last 50 years, since they show interesting phenomena such as a variety of magnetoelectric (ME) effects~\cite{Astrov60,Schmid94,Arima11}  and parity-mixed superconductivity~\cite{Bauer04}.
The intensive studies in the last decade have revealed that an antisymmetric spin-orbit coupling, which becomes active by space-inversion symmetry breaking, plays a relevant role in these phenomena.~\cite{Takahashi08,Cheong07,Frigeri04,Yanase14,Yoshida13}.
Furthermore, the very recent theoretical and experimental studies have revealed that the various ME effects can be better understood and categorized on the basis of spatially extended odd-parity multipoles, refered to as cluster or itinerant multipoles.~\cite{Nakatsuji15,Watanabe17,Suzuki17}.

A toroidal moment is the lowest-rank term of toroidal multipole tensors which appear in the multipolar expansion of an electromagnetic vector potential~\cite{Dubovik89}.
It can be active in the system without local space-inversion symmetry on the relevant ion sites. 
In a spin ordered system, the toroidal moment \textit{\textbf{t}} is defined as the summation of the vector products of position vector \textit{\textbf{r}}$_l$ and spin \textit{\textbf{S}}$_l$ for magnetic sites $l$: 
$\textit{\textbf{t}} = \frac{g\mu _{\rm{B}}}{2} \sum_{l}^{} \textit{\textbf{r}}_l \times \textit{\textbf{S}}_l$.
The summation is taken over appropriate magnetic basis.
In a system where toroidal moments order with a ferroic component, both time-reversal and global space-inversion symmetries are broken, and thus macroscopic ME effects can be expected to occur. 
For example, ME properties seen in high magnetic fields in a traditional multiferroic system Cr$_2$O$_3$~\cite{Popov99} and a novel nonreciprocal directional dichroism observed recently in LiCoPO$_4$~\cite{VanAken07} are described on the basis of the concept of toroidal order.
The toroidal moment has so far been discussed mainly in insulating systems, where \textit{\textbf{r}}$_l$ corresponds to an electric dipole (electric polarization), and the presence of {\boldmath$t$} in a system can easily be recognized.

Recently, Hayami {\it et al.} have theoretically investigated possible toroidal ordering in a metallic system with broken local-inversion symmetry at magnetic-ion sites~\cite{Hayami14}.
They predicted that exotic magnetotransport and ME effects can occur under the toroidal order.
Specifically, they performed a mean-field analysis for a single-band model on a layered honeycomb structure formed by one type of magnetic ion, and show that a ground state with the occurrence of spontaneous {\it toroidalization} {\boldmath$T$} (mean toroidal moment per unit volume) perpendicular to the layer planes is stabilized.
The most notable consequence in the theory will be the prediction of two types of ME response: one is net {\boldmath$T$} induced by an electric current perpendicular to the planes, which occurs even in paramagnetic state, and  the other is a uniform transverse magnetization induced by an electric current along the planes. 
Interestingly, they pointed out that an antiferromagnetic (AF) ordered state in UNi$_4$B corresponds to the ferroic toroidal order on a honeycomb structure, and thus may show the expected ME responses. 
In the present study, we have tested these theoretical predictions by measuring static magnetization of UNi$_4$B under applied electric currents. 
Note that the phenomenon of current-induced magnetization itself has quite recently been observed in a semiconductor, tellurium, even in the absence of any magnetic ordering~\cite{Furukawa17}.

\begin{figure}[t]
\begin{center}
\includegraphics[width=7cm]{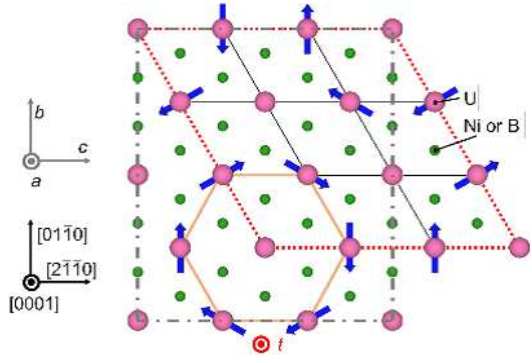}
\end{center}
\caption{(Color online) Schematic view of the reported crystal and magnetic structures of UNi$_4$B~\cite{Haga08,Mentink94,Mydosh92,Tabata}. Blue thick arrows indicate ordered magnetic moments of uranium ions (larger bullets) below $T$$\rm_N$~\cite{Mentink94}. A parallelogram (drawn with dotted lines) and hexagon (thick solid lines) denote an AF unit cell and a magnetic basis corresponding to a toroidal moment {\boldmath $t$}. Black (Gray) thin arrows indicate the crystal axes in the hexagonal (orthorhombic) notation, respectively. 
A rectangle (dash-dotted lines) corresponds to an orthorhombic unit cell in the a plane.
Ni and B occupy sites denoted by green bullets while reflecting each crystallographic symmetry, which have not been fully identified yet.}
\label{f1}
\end{figure}

It is currently considered that UNi$_4$B crystallizes in the orthorhombic structure with the symmetry {\it Cmcm} (No. 63, D$_{2h}$$^{\rm{17}}$).
Lattice parameters {\it a}, {\it b}, and {\it c} are 6.968 \AA, 17.1377 \AA, and 14.8882 \AA, respectively~\cite{Haga08}.
Once it had been reported to be the CeCo$_4$B-type hexagonal structure,~\cite{Mentink94} whereas recent high resolution~\cite{Haga08} and synchrotron X-ray~\cite{Tabata} studies revealed that the crystal has an orthorhombic structure which is formed by distorting the hexagonal structure and assigning different site occupation of Ni and B.
Figure 1 shows schematic illustration of hexagonal (0001) plane and magnetic structure.
The correspondence between the two crystal structures in the orthorhombic $a$ plane (the hexagonal (0001) plane) are also displayed.
As can be seen from the fact that estimated ratio $c$/$b$ $\sim$ 0.8687 is very close to the hexagonal value $\sqrt{3}/2$ ($\sim$ 0.8660), the distortion of a triangular lattice is so small that we could not distinguish between the $b$ and $c$ directions from the laboratory-based X-ray analysis.
Thus we use, hereafter, the hexagonal notation to describe the crystal directions and planes for simplicity.
Note that the actual U sites in orthorhombic structure are not located on the space-inversion center.

Neutron scattering studies show that UNi$_4$B exhibits an AF order at \textit{T}$\rm{_N}$ = 20.4 K, where only 2/3 of U ions participate the ordering.
The ordered magnetic moments lie in (0001) plane, forming a periodic array of vortex-like magnetic clusters in the shape of a hexagon (see Fig. 1)~\cite{Mydosh92}.
This magnetic cluster is equivalent to the definition of a toroidal moment.
Thus, we can describe the AF state of UNi$_4$B below \textit{T}$\rm{_N}$ as ferroic order of spontaneous toroidal moments pointing to [0001], if the reported magnetic structure is correct.
It is also reported that UNi$_4$B shows another phase transition at \textit{T}$^*$ = 0.3 K, where magnetic moments of the remaining 1/3 of U ions may order antiferromagnetically~\cite{Movshovich99}.


\begin{table}[b]
\label{t1}
\begin{center}
\caption{Summary of measurement conditions. Sample dimensions are stated in the order of the directions [2$\bar{1}$$\bar{1}$0], [01$\bar{1}$0], and [0001]. {\boldmath$B$}, {\boldmath$I$} and $i$ are applied magnetic field, electric current and electric current density, respectively.}
\begin{tabular}{cllll}
\hline
\multicolumn{1}{l} {Sample} &  &  &   \\
\multicolumn{1}{l} {Dimensions (mm$^3$)} & \multicolumn{1}{c}{Orientation} & \multicolumn{1}{c}{ $B$ (G)} & \multicolumn{1}{c}{$i$ (kA/m$^2$)}  \\
\hline
2.4 $\times$ 1.8 $\times$ 0.2 & \textit{\textbf{I}} \rm $||$ [2$\bar{1}$$\bar{1}$0], & 30 & 0, $\pm$27.8, $\pm$41.7, \\
                                        & \textit{\textbf{B}} \rm$||$ [01$\bar{1}$0]            &     &  $\pm$55.6, $\pm$69.4    \\
									        &										      								  & -5 & 0, $\pm$55.6 				   \\
0.2 $\times$ 2.4 $\times$ 3.1  & \textit{\textbf{I}} \rm $||$ [0001],                     & 30 & 0, 20.8, 41.7 					   \\
                                         & \textit{\textbf{B}} \rm$||$ [01$\bar{1}$0]           & 1   & 0, $\pm$41.7 			       \\
\hline
\end{tabular}
\end{center}
\end{table}

In the present work, the DC magnetization was measured using a commercial SQUID magnetometer (MPMS, Quantum Design Inc.) in the temperature range from 5 -- 50 K under magnetic field up to 30 G. 
A single-crystalline sample was grown by the Czochralski method using a tri-arc furnace, and confirmed to be a single phase from powder X-ray diffraction. 
No further heat treatment was performed.
The crystal was cut into rectangular parallelpiped shape with typical dimensions of $\sim$ 1.8 $\times$ 2.4 $\times$ 0.2 mm$^3$ using spark erosion.

In order to apply electric currents and check the electrical resistivity, four copper wires with a diameter of  0.026 mm  were introduced from the top of a sample probe intended for the DC-mode magnetization measurements.
The four wires were attached to the edges of the longer side of the samples using conductive silver paste.
The electric currents were applied using the DC current source, Model 6220 (Keithley Instruments Inc.).

The measurements were performed in the conditions of the electric currents {\boldmath $I$} parallel to [2$\bar{1}$$\bar{1}$0] and [0001], which are expected to correspond to the directions perpendicular and parallel to {\boldmath$T$}, respectively.
Two voltage terminals were used for checking {\boldmath$I$}  flowing through the samples. 
All the measurements were performed in the conditions of cooling under magnetic field {\boldmath$B$} (field cooling, FC) and a direct current (current cooling, CC). 
Measurement conditions are summarized in Table I.

\begin{figure}
\begin{center}
\includegraphics[width=7cm]{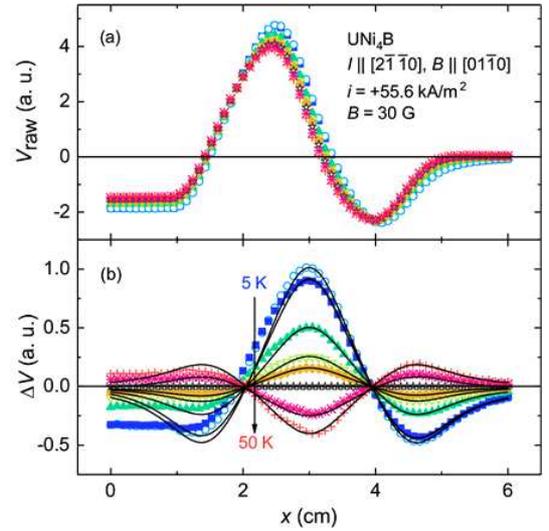}
\end{center}
\caption{(Color online) (a) Raw SQUID voltage $V_{\rm raw}$ and (b) its relative variations ${\it \Delta}V_{\rm raw}$ from the 30 K data versus position $x$. Solid curves indicate results of best fitting with the standard formula~\cite{MPMS}  for the fitting range 2.7 $< x <$ 6.0 cm.}
\label{f2}
\end{figure}

Figure 2(a) shows an example of raw voltage signals $V$ scaled by experimental factors as a function of position $x$.
The data are obtained by sweeping a sample through a second-derivative gradiometer (a set of pick-up coils) of the SQUID magnetometer, with applying {\it B} of 30 G and {\it I} of 20 mA to the sample.
This value of $I$ corresponds to the magnitude of electric current density, {\it i} = 55.6  kA/m$^2$.
In the standard measurements, the waveforms of $V(x)$ become symmetric, while those obtained under {\boldmath$I$} are obviously not.
This is mainly due to the additional magnetic flux generated by {\boldmath$I$} flowing through the sample and leads.
We assume that this background signal is independent of temperature, and subtract the $V_{0}(x)$ data taken at 30 K in the paramagnetic state from other temperature data for each $B$-$I$ condition.
The results of such a subtraction procedure for the data profile in Fig. 2(a) are given in Fig. 2(b).

The obtained relative variations of the output voltage, ${\it \Delta} V(x)$ = $V(x)$ $-$ $V_{0}(x)$, are still slightly asymmetric, showing a distortion from the standard formula in the position range below $\sim$ 2.5 cm.
This indicates that there is some unknown weak background that depends on temperature and cannot be subtracted only by using the data at 30 K.
Therefore, we use the ${\it \Delta} V(x)$ waveforms in the range 2.7 $< x <$ 6.0 cm for evaluating the magnetization $M$ in the present analyses.
 The solid curves in Fig. 2(b) indicate the results of the best fitting obtained by using the standard formula for $V(x)$.
All the data of $M$ presented below were obtained in the same manner.
Note that $M$ obtained by this procedure is the relative change  from that at 30 K. 

\begin{figure}[t]
\begin{center}
\includegraphics[width=7cm]{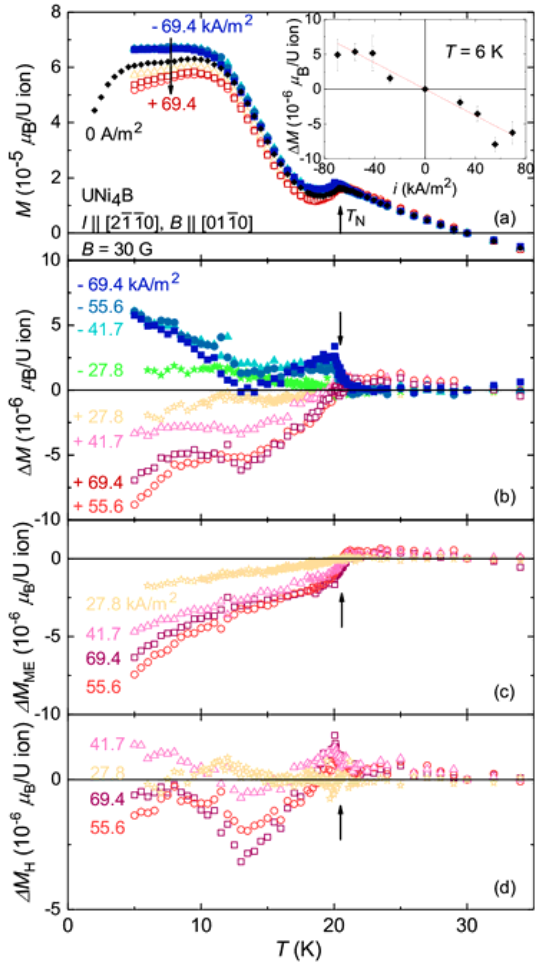}
\end{center}
 \caption{(Color online) Temperature dependence of (a) magnetization $M$, (b) current induced part of magnetization ${\it \Delta}M$, and contributions of (c) ME effects ${\it \Delta}M \rm_{ME}$ and (d) Joule heating ${\it \Delta}M \rm_{H}$ for various magnitudes of electric current density {\boldmath $i$}  ($||$ [2$\bar{1}$$\bar{1}$0]). The inset of (a) shows ${\it \Delta}M$ versus {\boldmath $i$} ($||$ [2$\bar{1}$$\bar{1}$0]) for $T$ = 6 K. The red line is the result of linear regression.}
\label{f3}
\end{figure}

Figure 3(a) shows the temperature dependence of {\it M} measured at 30 G in the direction of [01$\bar{1}$0] for $i$ = 0, $\pm$27.8, $\pm$41.7, $\pm$55.6, and $\pm$69.4 kA/m$^2$ applied along [2$\bar{1}$$\bar{1}$0].
We found that negative (positive) electric currents parallel to [2$\bar{1}$$\bar{1}$0] cause positive (negative) changes of $M$ in the direction of [01$\bar{1}$0] below around \textit{T}$\rm{_N}$, respectively.
By subtracting the data at 0 mA from those obtained under {\boldmath$I$}, the net component of magnetization, ${\it \Delta}M$, induced by {\boldmath$I$} is obtained as shown in Fig. 3(b).
${\it \Delta}M$ is almost constant (nearly zero) in the paramagnetic state for all the $i$ values, while the absolute magnitude of ${\it \Delta}M$ increases significantly as the temperature is lowered below \textit{T}$\rm{_N}$.


\begin{figure}[t]
\begin{center}
\includegraphics[width=7cm]{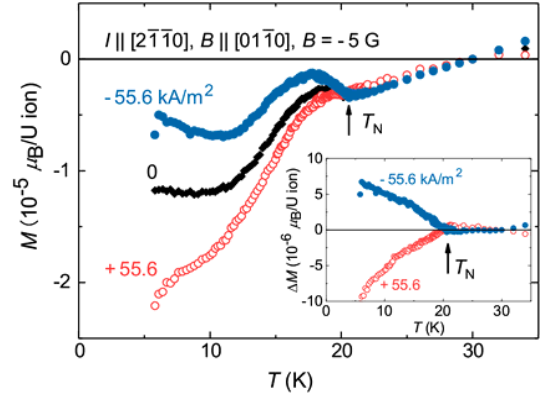}
\end{center}
\caption{(Color online) $M$ and ${\it \Delta}M$ (the inset) versus $T$ for $i$ = $\pm$ 55.6 kA/m$^2$ ($||$ [2$\bar{1}$$\bar{1}$0]) and $B$ = -5 G ($||$  [01$\bar{1}$0]).}
\label{f4}
\end{figure}

The inset of Fig. 3(a) shows the values of ${\it \Delta}M$ for {\boldmath $I$} $||$ [21$\bar{1}$0] at 6 K and 30 G as a function of {\it i}.
It is obvious that ${\it \Delta}M$ is in proportion to $i$ within the experimental accuracy.
The rate of increase d(${\it \Delta}M$)/d$i$ is estimated to be $\sim$ 9.4 $\times$ 10$^{-11}$ $\mu_B$m$^2$/(A $\cdot$ U).
We would like to emphasize again that the sign of ${\it \Delta}M$ is altered by reversing the direction of {\boldmath $I$}.
Since the theory predicts that ${\it \Delta}M$ occurs essentially in zero magnetic field~\cite{Hayami14}, we performed the same measurements at around zero magnetic field.
Figure 4 shows the data of $M$ taken in the same $B$-$I$ geometry for $i$ = 0 and $\pm$55.6 kA/m$^2$ at a weaker field of $-5$ G, which is a remanent field of the MPMS superconducting magnet in this run.
We observed that the obtained ${\it \Delta}M$ shows essentially the same behavior as that for $B$ = 30 G (the inset of Fig. 4).
The $|{\it \Delta}M|$ value at $\sim$ 6 K for $i$ = $\pm$55.6 kA/m$^2$ is estimated to be $\sim 7 \times 10^{-6 } \mu_B$/U, which is almost the same magnitude as that for $B$ = 30 G. The observed phenomenon is thus independent of the magnitude and the sign of {\boldmath$B$}, consistent with the theoretical prediction.

As seen in Fig. 3(b), all the ${\it \Delta}M$ curves are weakly winding in the temperature range 11 -- 18 K.
This behavior comes mainly from an extrinsic effect, ${\it \Delta}M$$\rm{_H}$($I^2$), caused by Joule heating on the sample, together with a strong temperature variation of $M$ in this temperature range.
The intrinsic contribution from the ME effects, ${\it \Delta}M$$\rm_{ME}$($I$), can be estimated as follows: ${\it \Delta}M$ should be described as
\begin{equation}
{\it \Delta} M_\pm = \pm {\it \Delta}M{\rm_{ME}}(I) + {\it \Delta}M{\rm_H}(I^2) \pm {\it \Delta}M{\rm_C}(I)
\end{equation}
(double-sign corresponds), where ${\it \Delta}M \rm_+$ and ${\it \Delta}M \rm_-$ indicate ${\it \Delta}M$ measured for $I$ $>$ 0 and $I$ $<$ 0, respectively, and ${\it \Delta}M {\rm_C} (I)$ denotes a residual contribution from magnetic flux generated by the introduced electric circuit, which can be negligible since this effect has already been subtracted during the fitting sequence of raw SQUID signal as mentioned.
Thus, ${\it \Delta}M \rm_{ME}$ and ${\it \Delta}M \rm_{H}$ can be calculated, respectively, as ${\it \Delta}M \rm_{ME}$ $\sim$ (${\it \Delta}M \rm_+$ $-$ ${\it \Delta}M \rm_-$)/2 and ${\it \Delta}M \rm_{H}$ $\sim$ (${\it \Delta}M \rm_+$ $+$ ${\it \Delta}M \rm_-$)/2, as shown in Figs. 3(c) and 3(d), respectively.
The analysis results may reveal the intrinsic behavior of ${\it \Delta}M \rm_{ME}$ and represent a significance of the heating effect around 13 K.

One might think that the assumption of ${\it \Delta}M \rm_{C}$ $\sim$ 0 is inappropriate; i.e., the contribution of a current loop cannot fully be eliminated by the subtraction of 30 K data.
Since the present measurements were performed under the constant {\boldmath$I$}, such an effect should be independent of a change in the electrical resistivity of the circuit composed of the sample and the Cu wires.
The only possible cause would be a change in the shape of the circuit due to thermal expansion.
However, a relative change in the sample length below 50 K is negligibly small ($\sim 10^{-6}$ K$^{-1}$)~\cite{Mentink97}, and thermal expansion of the Cu wires does not explain the fact that a significant change in ${\it \Delta}M$ occurs at \textit{T}$\rm{_N}$.

We would also like to mention a possible surface effect.
Since the Rashba-type antisymmetric spin-orbit coupling is always active on the surface of a metal, the observed ME phenomenon might be ascribed to a property of the surface states.
To test this possibility,  we repeated the measurements using the same sample piece after oxidizing the surface in the air;
the oxide thickness was estimated to be $\sim$ 0.03 mm by comparing the sample thickness before and after etching. 
Since the sequence of interfaces changes from metal-vacuum to metal-oxide-vacuum, the oxidization of the surface should affect the strength of the spin-orbit coupling, resulting in a change in the ME effects.
However, we observed no significant difference in the behavior of ${\it \Delta}M$ between the measurements with oxidized and non-oxidized surfaces.
We therefore naively suggest that the observed current-induced magnetization in UNi$_4$B is not simply attributed to the surface effect.
On the basis of these experimental results and consideration, we conclude that the change observed in ${\it \Delta}M$ parallel to [01$\bar{1}$0] below \textit{T}$\rm{_N}$ is intrinsic to the application of {\boldmath $I$} along the [2$\bar{1}$$\bar{1}$0] direction.
This is consistent with the theoretical prediction that uniform {\boldmath$M$} can be induced in the direction of {\boldmath $T$} $\times$ {\boldmath $I$}.


\begin{figure}[b]
\begin{center}
\includegraphics[width=7cm]{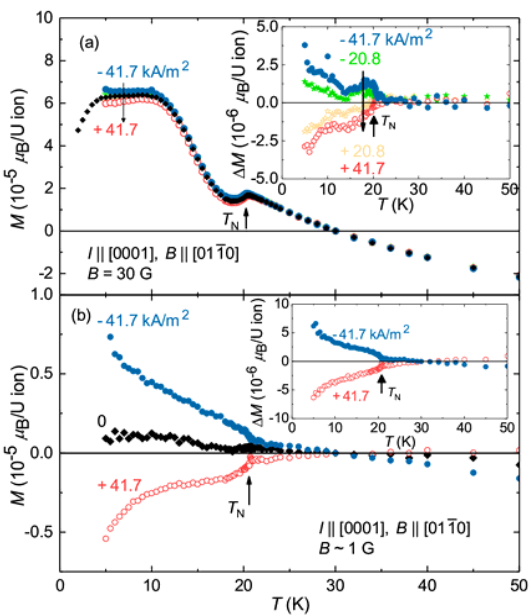}
\end{center}
\caption{(Color online) $M$ and ${\it \Delta}M$ (the insets) versus $T$ (a) for $i$ = $\pm$ 20.8 and $\pm$ 41.7 kA/m$^2$ ($||$ [0001]) and $B$ = 30 G ($||$  [01$\bar{1}$0]), and (b) for $i$ = $\pm$ 41.7 kA/m$^2$ ($||$ [0001]) and $B$ $\sim$ 1 G ($||$  [01$\bar{1}$0])}
\label{f5}
\end{figure}

Figure 5(a) shows the experimental results obtained in the different geometry: $M$ $||$ [01$\bar{1}$0] under {\boldmath $I$} $||$ [0001] at $B$ = 30 G. 
Obviously, $M(T)$ is enhanced or suppressed below and near \textit{T}$\rm{_N}$ by applying {\boldmath$I$} in positive or negative directions, respectively.
The obtained ${\it \Delta}M$ stays nearly constant in the paramagnetic state, while changes as the temperature is lowered below \textit{T}$\rm{_N}$ (the inset of Fig. 5(a)), roughly in proportion to $i$.
The winding feature with a shallow minimum at $\sim$ 13 K of ${\it \Delta}M$ is considered to be due to the Joule-heating effects on the sample, as mentioned above.

Figure 5(b) shows the experimental results obtained in the same $B$-$I$ geometry at a weaker $B$ of $\sim$ 1 G (again, the remanent field in this run).
The reduction of $B$ makes the temperature variation of ${\it \Delta}M$ clearer, obscuring the heating-up effects.
The absolute magnitude of ${\it \Delta}M$ at 6 K and $\pm$41.7 kA/m$^2$ is estimated to be about $5 \times 10^{-6 } \mu_B$/U, which is slightly smaller than that for $M$ $||$ [01$\bar{1}$0] and {\boldmath $I$} $||$ [2$\bar{1}$$\bar{1}$0].

According to the theory proposed by Hayami \textit{et al.}~\cite{Hayami14},  {\boldmath $I$} $||$ [0001] of UNi$_4$B may induce the change of toroidalization, as follows. 
Below $T_{\rm N}$, the ferroic order of {\boldmath $t$} should form two domains with {\boldmath $T$} $||$ [0001] or [000$\bar{1}$]. In each domain of {\boldmath $T$} $||$ $+$ {\boldmath $I$} or $-$ {\boldmath $I$}, {\boldmath $T$} will be enhanced or suppressed with {\boldmath $I$}, where microscopically the ordered magnetic moments are enlarged or reduced, respectively. 
In addition, the electric currents {\boldmath $I$} $||$ [0001] ([000$\bar{1}$]) may flip the magnetic moments in the domain with {\boldmath $T$} $||$ [000$\bar{1}$] ([0001]), and thus change the volume fraction of two domains. 
Such changes in magnetic moments are expected to be detected as a change in $M$ in magnetic fields applied in the basal (0001) plane, if two domains have different volumes. 
Moreover, the theory predicts that {\boldmath $I$} induces {\boldmath $T$} even in the paramagnetic state, which will suppress $M$, independently of the direction of {\boldmath $I$}.
In the present measurements for {\boldmath $I$} $||$ [0001], however,  
${\it \Delta}M$ is observed only below $T_{\rm N}$ and its magnitude is almost independent of {\it B}.
This behavior is significantly different from the theoretical prediction.

One reason of the above inconsistency could be a path that {\boldmath$I$} flow through in a sample.
In the measurements for {\boldmath $I$} $||$ [0001], we used a sample piece which has a rectangular parallelpiped shape with a width of $\sim$ 2.4 mm and a thickness of 0.2 mm  perpendicular to the current flow direction.
Therefore, the currents that flow through the sample may have a component in the basal (0001) plane, which could result in the ${\it \Delta}M$ as we observed for {\boldmath $I$} $||$ [2$\bar{1}$$\bar{1}$0].
In order to check this possibility, it is necessary to repeat the measurements using a sample with different dimensions.
Our preliminary repeated trials, however, show no significant change in the behavior of ${\it \Delta}M$ thus far.
We should note that the current path also depends on the anisotropy of the electrical resistivity $\rho$ in general.
The $\rho$ of UNi$_4$B for {\boldmath $I$} $||$ [2$\bar{1}$$\bar{1}$0] is several times larger than that for {\boldmath $I$} $||$ [0001]~\cite{Mentink94}.
Therefore, from the viewpoint of anisotropy in $\rho$, the deviation of current flow from the [0001] direction will be unfavorable.

Another, more likely reason is derived from the ambiguity of the crystal and magnetic structures of the present system.
Mentink {\it et al.} proposed vortex-like magnetic structure below \textit{T}$\rm{_N}$ on the basis of their neutron diffraction measurements~\cite{Mentink94}.
However, the analysis they made is based on the hexagonal CeCo$_4$B-type crystal structure, and the best fitting obtained among a few hundred candidates still has a large reliability factor \textit{R} = 11.8\%.
In addition, if the crystal structure is orthorhombic, {\it Cmcm}~\cite{Haga08,Tabata}, the arrangement of Ni and B atoms surrounding U atoms differs largely from that in the hexagonal structure: the former has four inequivalent crystallographic U sites in a U-Ni-B layer, while the latter has two U sites forming U-Ni and U-B layer. 
The four U sites in the orthorhombic structure do not have the inversion symmetry, and thus may produce local toroidal moments which are different both in direction and magnitude at each U site.
On the other hand, the theory assumes that an odd-parity crystalline electric field exists along radial direction from the center to the vertices of alternate uranium hexagons, and local toroidal moments emerge identically on each vertex.
This inconsistency between the theory and the present experimental results thus implies that the crystal and magnetic structures differ from those discussed so far.

In summary, we performed magnetization measurements under electric currents in the AF uranium compound UNi$_4$B.
We have revealed that the application of electric currents parallel to [2$\bar{1}$$\bar{1}$0] and [0001] both induces static magnetization of the order of 10$^{-10} \mu_\text{B}$/U per unit current density, in the direction of [01$\bar{1}$0] below \textit{T}$\rm{_N}$.
The observation is consistent with the recent theoretical predictions, in the sense that the ME effects may actually occur in a metallic system with broken local-inversion symmetry.
However, we have also observed a crucial inconsistency that ${\it \Delta}M$ is induced by the current flow {\boldmath $I$} $||$ [0001], which is an inactive geometry, {\boldmath $I$} $||$ {\boldmath $T$}, regarding the ME effects.
In order to gain a better understanding of the origin of the observed phenomena, we need to complete the identification of crystal and magnetic structures of this system.
Theoretical verification of the observed magnitude of the ME effects will also be a crucial issue for future studies.

\begin{acknowledgment}
The authors thank S. Hayami, Y. Motome, H. Kusunose, H. Tou, H. Harima, M. Vali\v{s}ka, K. Uhl\'{i}\v{r}ov\'{a}, and V. Sechovsk\'{y} for fruitful discussions.
The present research was supported by JSPS Grants-in-Aid for Scientific Research (KAKENHI) Grant No. JP15K1350905, and Grant Nos. JP15H05882 and JP15H05885 (J-Physics).
\end{acknowledgment}


\end{document}